\def \cmsq           {\hbox{cm$^{-2}$}}
\def \etal         {{\it et~al.} }

\def \kms          {\rm{\hbox{km s$^{-1}$}}}
\def \lam          {$\lambda$}
\def \Lya          {\hbox{Ly$\alpha$}}

\def \mum          {\hbox{$\mu$m}}

\def \zaz          {{$z_a\kern -1.5pt \approx\kern -1.5pt z_e$}}
\def \zllz         {{$z_a\kern -3pt \ll\kern -3pt z_e$}}

\def \zgz          {{\kiA z\lower 3pt \hbox{a} $>$ z\lower 3pt \hbox{e}\ }}
\def \Msun         {\rm{\hbox{M$_{\odot}$}}}           
\def \Zsun         {\rm{\hbox{Z$_{\odot}$}}}           
\documentstyle[11pt,paspconf]{article}

\begin{document}

\title{Ionization and Abundances in Intrinsic QSO Absorption-Line Systems}
\author{F. Hamann, T. Barlow, R.D. Cohen, 
V. Junkkarinen \& E.M. Burbidge}
\affil{Center for Astrophysics \& Space Sciences, University of California 
-- San Diego, La Jolla, CA, 92093-0424}

\begin{abstract}
We discuss two aspects of the ionization and abundances in 
QSO intrinsic absorbers. First, we use the high-ionization 
doublet Ne~{\sc viii}~\lam\lam 770,780 to test the relationship between 
the UV line absorbers and the ``warm'' absorbers detected in soft X-rays. 
In one well-measured QSO with intrinsic \zaz\ lines, UM 675, 
we estimate that the UV absorber has at least 10 times too little 
O~{\sc vii} and O~{\sc viii} to produce the requisite bound-free edges at 
$\sim$0.8 keV. Second, we show that firm lower limits on the metal-to-hydrogen 
abundances can be established even with no constraints on the 
ionization. Those lower limits are typically $Z\ga$~\Zsun\ and sometimes 
$Z\ga 10$~\Zsun\ (e.g. for the BALs). We argue that QSO metallicities 
up to at least $\sim$9~\Zsun\ are consistent with the rapid early-epoch 
star formation expected in the cores of massive galaxies. 
\end{abstract}


\section{General Introduction}

Intrinsic QSO absorption lines include the broad absorption lines (BALs), 
at least some of the much narrower associated (\zaz ) lines, and 
any other (\zllz ) systems that form near the QSO engine. 
We are involved in a program to identify these systems and 
study their kinematics, 
ionization and metal abundances using spectra from 
the {\it Hubble Space Telescope} ({\it HST}) and the Keck and Lick 
Observatories. Some results on the identifications and 
kinematics are discussed in the accompanying paper by Hamann \etal 
in this volume (hereafter Paper1).

\section{Ne~{\sc viii} and the UV Line--X-ray Warm Absorber Connection}

The combined UV ({\it HST}) and visible (ground-based) spectra allow 
us to (1) measure lines across a wide range of ionizations and (2) 
derive reliable column densities from the relatively low-resolution 
HST spectra. 
We are particularly interested in the O~{\sc vi}~\lam\lam 1032,1037 and 
Ne~{\sc viii}~\lam\lam 770,780 doublets to probe the high-ionization gas 
and test the possible relationship (Mathur \etal 1994; also Mathur 
\etal and Shields \& Hamann this volume) 
between the UV-line and X-ray warm absorbing regions. Warm absorbers 
can be identified by bound-free absorption edges 
of O~{\sc vii} and/or O~{\sc viii} at $\sim$0.8 keV. Measurements of the 
Ne~{\sc viii} column density can directly constrain the amount of warm 
absorber material because O~{\sc vii} and O~{\sc viii} are the dominant 
species of oxygen in the Ne~{\sc viii} absorbing gas (see below).

We now have two firm detections of the Ne~{\sc viii} doublet.   
Both are in \zaz\ systems of redshift $\sim$2 QSOs: PKS~0119$-$046 
and UM~675. All of the strong \zaz\ lines in PKS~0119$-$046, including 
the Ne~{\sc viii} pair, appear to be saturated and therefore do not provide 
accurate column densities. 
The results for UM~675 are more reliable. Figure 1 shows 
the {\it HST}-FOS spectrum of this source (Junkkarinen \etal 1997). 
Much higher resolution Keck Observatory spectra 
across the \Lya , C~{\sc iv} and N~{\sc v} lines are shown in Paper1 
and Hamann \etal (1997). Table 1 lists the 
derived column densities ($\log N$ in \cmsq ) 
for each of the \zaz\ lines. The column densities are from Hamann \etal 
(1995) except that the metallic columns are increased by a factor 
of 2 to be consistent with the 50\% coverage fraction derived for the 
C~{\sc iv} and N~{\sc v} absorbers (Paper1; Hamann \etal 1997). Note that 
this scaling is just a crude correction because different ions can 
cover different fractions of the background light source(s) 
(Hamann \etal 1997; Barlow \& Sargent 1997). 

\begin{figure}
\plotfiddle{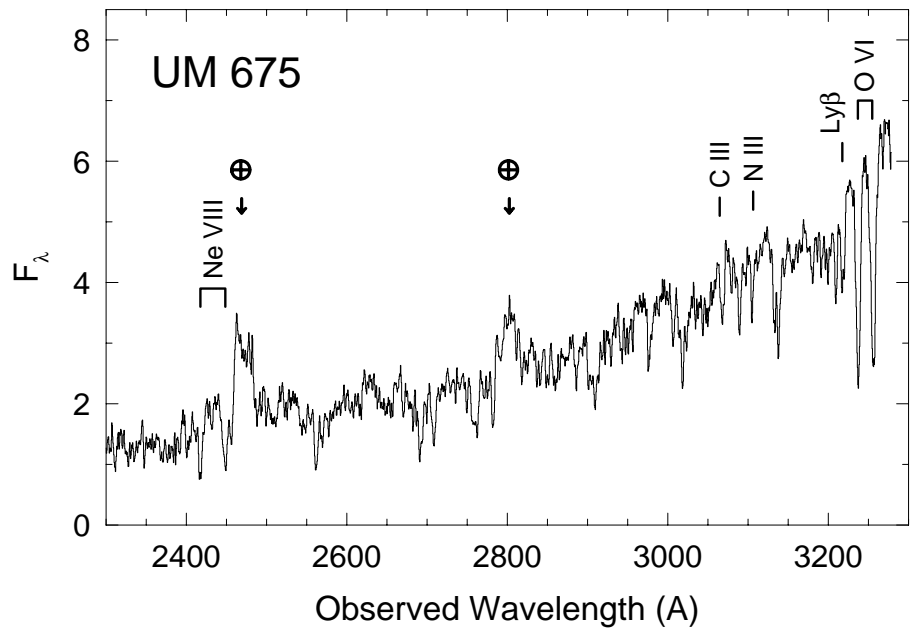}{3.1in}{0.0}{60.}{60.}{-243.0}{-55.0}
\caption{Pre-COSTAR {\it HST}-FOS spectrum of UM~675 at 
$\sim$230~\kms\ resolution. Most of the absorption lines 
are due to the \Lya\ forest. The strong \zaz\ lines are labelled. 
The two marked emission features are terrestrial airglow.} 
\end{figure}

\begin{table}
\caption{Example Column Densities and Abundances} \label{tbl-1}
\begin{center}\scriptsize
\begin{tabular}{lcccc}
Ion& log $N$(M$_i$)& & [M/H]$_p$ & [M/H]$_{min}$\\
\noalign{\vskip 2pt}
\tableline
\noalign{\vskip 4pt}
\multicolumn{5}{c}{UM 675 ($z_a\approx z_e\approx 2.14$)}\\
\noalign{\vskip 2pt}
\tableline
\noalign{\vskip 4pt}
H~I& 14.8&     & ---& ---\cr
\noalign{\vskip 2pt}
C~III& 14.1&     & $+$0.4$^{+0.5}_{-0.2}$ & $-$0.2$^{+0.7}_{-0.2}$\cr
\noalign{\vskip 2pt}
C~IV& 14.9&    & $+$0.4$^{+0.7}_{-0.3}$ & $-$0.1$^{+0.9}_{-0.3}$\cr
\noalign{\vskip 2pt}
N~III& 14.5&    & $+$1.3$^{+0.5}_{-0.2}$ & $+$0.7$^{+0.7}_{-0.3}$\cr
\noalign{\vskip 2pt}
N~V& 15.1&    & $+$0.3$^{+0.8}_{-0.4}$ & $-$0.2$^{+0.9}_{-0.3}$\cr
\noalign{\vskip 2pt}
O~VI& 15.5&    & $-$0.9$^{+0.8}_{-0.3}$ & $-$1.4$^{+0.8}_{-0.3}$\cr
\noalign{\vskip 2pt}
Ne~VIII& 15.7&  & $-$1.1$^{+0.7}_{-0.4}$ & $-$1.4$^{+0.7}_{-0.4}$\cr
\noalign{\vskip 3pt}
\tableline
\noalign{\vskip 5pt}
\multicolumn{5}{c}{Mean BAL ($z_e\approx 2$)}\\
\noalign{\vskip 2pt}
\tableline
\noalign{\vskip 4pt}
H~I& 15.4&         & ---& ---\cr
\noalign{\vskip 1pt}
C~IV& 16.0& & $+1.0^{+0.7}_{-0.4}$& $+0.5^{+0.8}_{-0.4}$\cr
\noalign {\vskip 2pt}
N~V& 16.1& & $+0.7^{+0.8}_{-0.4}$& $+0.5^{+0.8}_{-0.4}$\cr
\noalign {\vskip 2pt}
O~VI& 16.3\rlap{:}& & $-0.6^{+0.7}_{-0.3}$& $-1.1^{+0.7}_{-0.4}$\cr
\noalign {\vskip 2pt}
Si~IV& 15.2& & $+1.6^{+0.6}_{-0.4}$& $+1.4^{+0.6}_{-0.4}$\cr
\noalign {\vskip 2pt}
\tableline
\end{tabular}
\end{center}

\end{table}

\begin{figure}
\plotfiddle{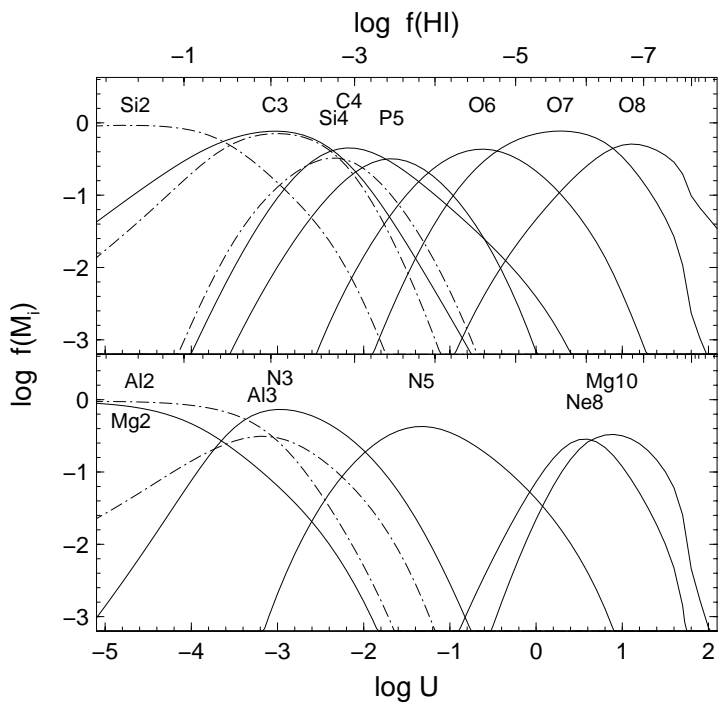}{3.5in}{0.0}{58.}{58.}{-170.0}{-167.0}
\caption{Ionization fractions in photoionized, optically thin clouds. 
The curves for each metal ion are labeled at their peaks, except for 
Si3, which is the dash-dot curve between Si2 and Si4 in the top panel. 
The notation is Si3~=~Si~{\sc iii}~=~Si$^{+2}$, etc.} 
\end{figure}

Figure 2 plots theoretical ionization fractions, $f$(M$_i)$, for 
various metals, M, in ion stage, $i$, as a function of the ionization 
parameter, $U$, in photoionized, optically thin clouds 
(where $U$ is the dimensionless ratio of hydrogen-ionizing 
photon to hydrogen particle densities at the illuminated face of the 
clouds). The H~{\sc i} fraction, $f$(H~{\sc i}), is shown across the top of 
the figure. The calculations were performed using CLOUDY (version 
90.02, Ferland 1997) with an ionizing spectrum that is believed to be 
typical of QSOs and Seyfert 1 nuclei (from Mathews \& Ferland 1987, 
with an additional decline at wavelengths $\geq$2 \mum ). 
The results do not depend on the column densities  
as long as the clouds are optically thin in the H~{\sc i} Lyman 
continuum, as is implied by the measured column densities 
in most \zaz\ and BAL systems -- including UM~675 (Hamann 1997). 

Comparing the measurements in Table 1 to the calculations 
in Figure 2 shows that there is a range of 
ionization states in the UM~675 absorber. The column density ratios 
of C~{\sc iii}/C~{\sc iv} and N~{\sc iii}/N~{\sc v} yield an 
abundance-independent estimate of $U\approx 0.02$, while the ratio 
O~{\sc vi}/Ne~{\sc viii} implies $U\approx 5$ (with the reasonable 
assumption that the O/Ne abundance ratio is solar). 
This range of ionization states requires that the clouds span 
factors of $\sim$250 in density, $\sim$16 
in distance from the ionizing QSO, or some equivalent 
combination thereof. 

Figure 2 also shows that we can estimate the column density 
in O~{\sc vii}+O~{\sc viii} from $N$(Ne~{\sc viii}) by the relation, 
\begin{equation}
N({\rm OVII+OVIII}) \ \approx \ \ N({\rm NeVIII}) 
\left({{\rm Ne}\over{\rm O}}\right)
{{f({\rm OVII+OVIII})}\over{f({\rm NeVIII})}}
\end{equation}
where Ne/O is the abundance ratio and 
$f$(Ne~{\sc viii})/$f$(O~{\sc vii}+O~{\sc viii}) is an ionization 
correction. For solar Ne/O and a conservatively large ionization correction 
factor of 10 (Fig. 2), 
we find $N$(O~{\sc vii}+O~{\sc viii})~$\approx 4\times 10^{17}$~\cmsq , 
which implies bound-free optical depths at the O~{\sc vii} and O~{\sc viii} 
edges of 0.08 and 0.04, respectively. We conclude that there is highly 
ionized warm-absorber-like gas in the UV line forming region of UM~675, 
but there is not enough of that gas to actually be a warm absorber. 
With H. Netzer and J. Shields, we are now pursuing simultaneous 
observations of the Ne~{\sc viii} lines and O~{\sc vii}+O~{\sc viii} 
edges in other QSOs to test the UV line--X-ray warm absorber connection 
further.

\section{Metal Abundances and Host Galaxy Evolution}

The normalized abundance of any metal relative to hydrogen is given by, 
\begin{equation}
\left[{\rm M\over H}\right] \ = \ \ 
\log\left({{N({\rm M}_i)}\over{N({\rm HI})}}\right) \ +\
\log\left({{f({\rm HI})}\over{f({\rm M}_i)}}\right) \ +\
\log\left({{{\rm H}}\over{\rm M}}\right)_{\odot}
\end{equation}
where (H/M)$_{\odot}$ is the solar abundance ratio. 
If the gas is in photoionization equilibrium 
and optically thin at all far-UV continuum wavelengths, the 
ionization correction factors, $f$(H{\sc i})/$f$(M$_i$),  depend only on 
$U$ and the shape of the ionizing spectrum. Column density 
measurements for a variety of ions can constrain $U$ and the correction 
factors needed for [M/H] by comparison to Figure 2. Further comparisons 
to calculations using other spectral shapes can define the uncertainties 
(see Hamann 1997). 

For systems with a range of ionization states  
(e.g. UM~675) or no useful constraints on $U$, we can derive 
conservatively low values of $f$(H{\sc i})/$f$(M$_i$) 
and therefore [M/H] by assuming that each 
metal line forms where that ion is most abundant, i.e. at an ionization 
corresponding to the peak of its $f$(M$_i$) curve in Figure 2. 
We can also place firm lower limits on the [M/H] ratios by adopting  
the minimum ionization corrections for each M$_i$. 
Every correction factor reaches a  
minimum value at $U$ values slightly larger than the peak in the 
$f$(M$_i$) curve. Figure 3 shows two examples of the 
minimum correction factors for a range of ionizing spectral shapes 
(from Hamann 1997). 
The spectra used for these calculations are segmented power laws with fixed 
slopes ($f_{\nu}\propto \nu^{\alpha}$) of $\alpha = -0.5$ in the visible and 
$-$1.0 in X-rays above 0.7 keV. We explore a range of continuum shapes 
by varying $\alpha_{ox}$, which relates the flux densities 
($f_{\nu}$) at 2500~\AA\ and 2~keV, and $E_c$, which marks the cutoff 
energy where the visible power law gives way to a steeper decline toward 
the X-ray flux at 0.7 keV. 
(This is continuum `C' in Hamann 1997.)
The Mathews \& Ferland (1987) continuum used in Figure 2 can be 
approximated by $\alpha_{ox}\approx -1.5$ and 
$\log E_c$~(eV)~$\approx 1.5$.  

\begin{figure}
\plotfiddle{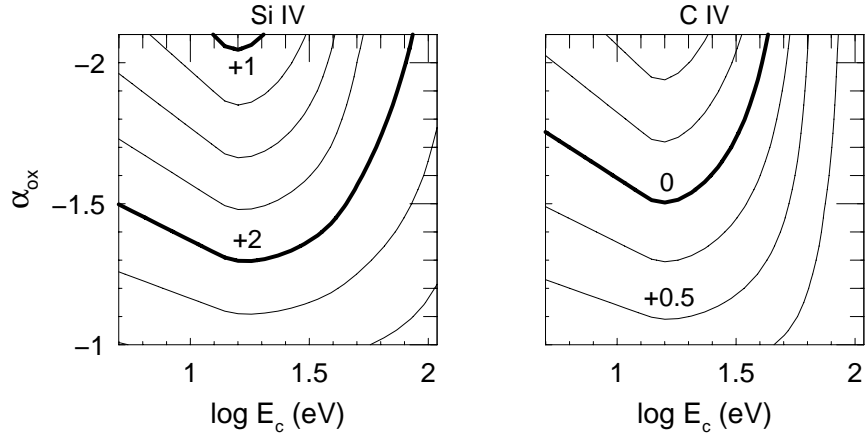}{2.1in}{0.0}{65.}{65.}{-200.0}{-230.0}
\caption{Contours of constant minimum ionization correction 
normalized by solar abundances, 
log($f$(H~{\sc i})/$f$(M$_i$))~+~log(H/M)$_{\odot}$, are plotted for Si IV 
and C IV in optically thin clouds 
photoionized by a range of QSO continua. The parameters 
$\alpha_{ox}$ and $E_c$ define the continuum shapes. 
Bold contours appear every 1.0 dex and thin contours every 0.25 dex.} 
\end{figure}

The last two columns in Table 1 apply these calculations 
to the \zaz\ absorber in UM~675 and to a set of mean column 
densities derived for BALs (Hamann 1997). [M/H]$_p$ are the 
conservatively low abundance ratios that follow from the assumption 
that each metal line forms at the peak in its $f$(M$_i$) curve. 
[M/H]$_{min}$ are the lower limits derived from the minimum 
correction factors in Figure 3 (also Hamann 1997). 
The uncertainties listed in the table indicate the range of values 
derived for a range of continuum shapes. 
Both [M/H] quantities are typically higher (and more 
realistic) for low-ionization metals because the H~{\sc i} lines 
tend to form with these ions. The [M/H]$_p$ results provide our best 
guess at the actual abundances in UM~675;  
the overall metallicity is roughly twice  
solar ($Z\approx 2$~\Zsun ) based on [C/H], with nitrogen 
several times more enhanced. The average BAL column densities 
indicate that they typically have much higher metallicities of 
$Z\ga 25$~\Zsun\   
based on [Si/H]. However, that result and the extremely high 
P/C ratios reported for some BALs (Turnshek \etal 1996 and this volume; 
Korista \etal 1996; Junkkarinen \etal this volume) are 
uncertain because the column densities depend on the unknown 
coverage fractions of the absorbers (also Paper1 and Arav this volume). 
Nonetheless, the main result for $Z\ga$~\Zsun\ does appear secure and 
typical of all types of intrinsic absorbers (also Petitjean, Rauch \& 
Carswell 1994; Hamann 1997). There is also independent evidence 
for $Z\ga$~\Zsun\ in QSOs from the broad emission lines 
(Hamann \& Ferland 1993; Ferland \etal 1996).

This growing evidence for high abundances in QSOs supports standard 
models of galaxy evolution; vigorous star formation in the spheroidal 
cores of massive galaxies should produce super-solar gas-phase 
metallicities within a few billion years of the initial collapse. 
High metal abundances are a signature of {\it massive} 
galaxies because only they can retain their gas long enough 
against the building thermal pressures from supernova explosions. 
The enriched gas might ultimately be ejected from the 
galaxy, consumed by the black hole, or diluted by subsequent infall, 
but the evidence for early-epoch high-$Z$ gas remains in the stars today.  
In particular, the mean stellar metallicities in the cores of nearby 
massive galaxies are typically $\sim$1 to 3~\Zsun . The individual stars 
are distributed about these means with metallicities reflecting the 
gas-phase abundance at the time of their formation. Only the most 
recently formed stars at any epoch have metallicities as high as 
that in the gas. Standard chemical evolution models 
indicate that the gas-phase abundances in galactic nuclei are 
$\sim$2 to 3 times larger than the stellar means, e.g.   
$\sim$2 to 9~\Zsun\ near the end of the star-forming epoch 
(see Hamann \& Ferland 1993, Hamann 1997 and references therein). 
Therefore, metallicities in the range $2\la Z\la 9$~\Zsun\ 
can be {\it expected} in QSOs as long as (1) the gas we observe 
was processed by stars in the cores of massive ($\ga$10$^{11}$~\Msun ) 
galaxies (or at least in dense condensations that become the cores 
of massive galaxies), and (2) most of the star formation and enrichment 
occurs before the QSOs ``turn on'' or become observable.

\acknowledgments
This work was supported by NASA grants NAG 5-1630 and NAG 5-3234.

\end{document}